\documentclass[twocolumn,aps,prl,showpacs, amsmath,amssymb,floatfix]{revtex4}

\usepackage{amsmath}
\usepackage{amssymb}
\usepackage{graphicx}
\usepackage{bm}

\begin{document}

\title{Piecewise Interference and Stability of Branched Flow}

\author{Bo Liu}
\email{bliu@physics.harvard.edu}
\affiliation{Department of Physics, Harvard University, Cambridge, Massachusetts 02138, USA}

\date{\today}

\begin{abstract}
The defining feature of chaos is its hypersensitivity to small perturbations. However, we report a stability of branched flow against large perturbations where the classical trajectories are chaotic, showing that strong perturbations are largely ignored by the quantum dynamics. The origin of this stability is accounted for by the piecewise nature of the interference, which is largely ignored by the traditional theory of scattering. Incorporating it into our theory, we introduce the notion of piecewise classical stable paths(PCSPs). Our theory shall have implications for many different systems, from electron transport in nanostructures, light propagation in nonhomogeneous photonic structures to freak wave formations in oceans.
\end{abstract}

\pacs{72.10.-d, 05.45.Mt, 05.60.Gg}
\maketitle

Chaotic dynamics is usually characterized by its hypersensitivity to small perturbations. Even a slight change in initial conditions is exponentially magnified over time in such systems, popularly known as the "butterfly effect". This hypersensitivity is usually undesirable, but many groups have managed to take advantage of it\cite{a3,a4,a2} for applications in optics and controlling chaos. How classical chaos manifests itself in quantum systems is an interesting  question being actively studied\cite{a5,a6,a7,a8,e1,e2} and worth further exploration. In this Letter, we attempt to address the fundamental problem:"How should we visualize quantum dynamics when the classical dynamics is chaotic?"

In the following, we study the quantum dynamics in the  general case where we have an open quantum system assuming nothing except for classical chaotic trajectories that are exponentially unstable to perturbations. For simplicity, we restrict our discussion to two dimensions and generalization to higher dimensions is straightforward. 

All information about the quantum dynamics is encoded in the propagator $G(\vec{q}_{f},t_{f};\vec{q}_{0},t_{0})=<\vec{q}_{f},t_{f}|e^{-iH(t_{f}-t_{0})/\hbar}|\vec{q}_{0},t_{0}>$, which can be written in terms of the Feynman Path Integral as
\begin{equation} 
G(\vec{q}_{f},t_{f};\vec{q}_{0},t_{0})=A\sum_{j} e^{iS_{j}(\vec{q}_{f},t_{f};\vec{q}_{0},t_{0})}
\label{pp} \end{equation}
, where $S_{j}(\vec{q}_{f},t_{f};\vec{q}_{0},t_{0})=\int_{t_{0}}^{t_{f}} ds\ L(\vec{q}_{j}(s),\dot{\vec{q}}_{j}(s),s)$ is the action for the jth path and $L(\vec{q}_{j}(s),\dot{\vec{q}}_{j}(s),s)$ is the Lagrangian. The summation is over all possible paths connecting ($\vec{q}_{0},t_{0}$) and ($\vec{q}_{f}$,$t_{f}$) and A is a normalization constant.

One can approximate (\ref{pp}) by the Van Vleck-Gutzwiller(VVG) propagator in two dimensions\cite{b13}:
\begin{equation} 
G(\vec{q}_{f},t_{f};\vec{q}_{0},t_{0})\approx \frac{1}{2i\pi\hbar} \sum_{j} \left | \frac{\partial^{2} S_{j}}{\partial \vec{q}_{0} \partial \vec{q}_{f}}\right |^{\frac{1}{2}} e^{iS_{j}(\vec{q}_{f},t_{f};\vec{q}_{0},t_{0})/\hbar -iv_{j}\pi/2}
\label{vvg} \end{equation}
, where the summation is over only classical trajectories connecting ($\vec{q}_{0},t_{0}$) and ($\vec{q}_{f},t_{f}$) and the Maslov index $v_{j}$ increases by one whenever $\left | \frac{\partial^{2} S_{j}}{\partial \vec{q}_{0} \partial \vec{q}_{f}}\right |$ becomes singular. We can gain more insights by noting that
\begin{equation} 
\left |\frac{\partial^{2} S_{j}}{\partial \vec{q}_{0} \partial \vec{q}_{f}}\right |=\left |\frac{\partial \vec{p}_{0}}{\partial \vec{q}_{f}}\right |=\left |\frac{\partial \vec{q}_{f}}{\partial \vec{p}_{0}}\right |^{-1}
\label{prefactor} \end{equation}.

$\left |\frac{\partial \vec{q}_{f}}{\partial \vec{p}_{0}}\right |$ measures how much a small change in initial momentum is magnified over time in terms of the change in the final position. Intuitively, this quantifies how chaotic each individual trajectory is. We can now interpret (\ref{vvg}) as the following:  we send out classical trajectories with all possible momenta at ($\vec{q}_{0}$,$t_{0}$)  and then count the contributions from those that reach $(\vec{q}_{f}$,$t_{f}$) with more weights given to stable trajectories. 

However, VVG is based on the stationary phase approximation and we either need $\hbar \to 0$ or $t_{f}-t_{0}  \to 0$ for it to be accurate. The first condition is the classical limit, which is not what we are after. Instead, we divide time into small intervals. Define $t_{k}=t_{0}+k\tau$, $k=0,1,2,\cdots,N$, $\tau=\frac{t_{f}-t_{0}}{N}$  and rewrite the propagator as 
\begin{equation} 
\begin{aligned}
G(\vec{q}_{f},t_{f};\vec{q}_{0},t_{0})=&\frac{1}{(2i\pi\hbar)^{N}} \sum_{j} \prod_{k=0}^{N-1} \left |\frac{\partial \vec{q}_{k+1}}{\partial \vec{p}_{k}}\right |^{-\frac{1}{2}} \\
&\times e^{iS_{j}(\vec{q}_{k+1},\vec{q}_{k},t_{k+1},t_{k})/\hbar -iv_{j,k,k+1}\pi/2}
\end{aligned}
\label{fp} \end{equation}

\begin{figure*}
\begin{center}
\includegraphics[width=6.5in]{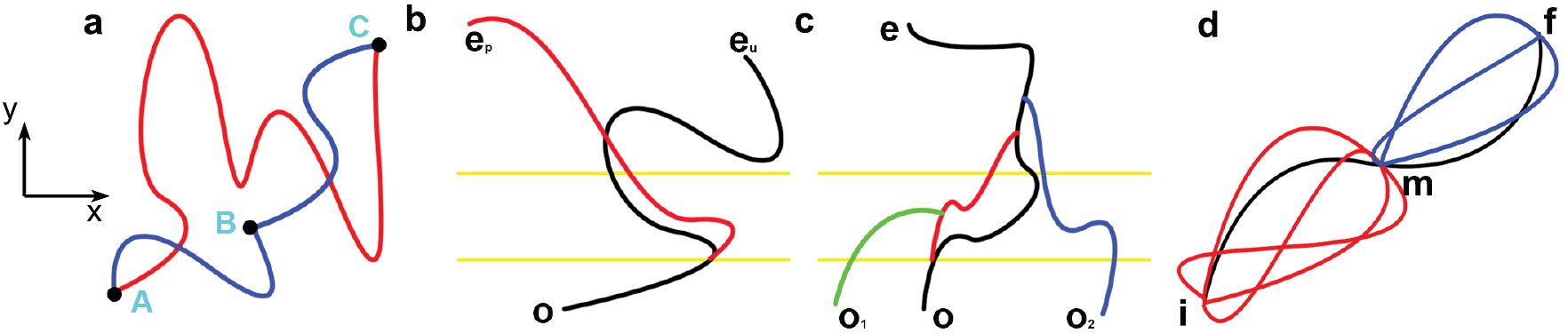}
\end{center}
\caption{(color online)\label{path}\footnotesize{\textbf{Classical Dynamics vs. Quantum Dynamics.}  ({\bf a}) The red trajectory is one classical trajectory connecting A and C in a period of $\tau$, while the blue trajectory is a PCT containing two classical trajectories from A to B in $\tau/2$ and from B to C in $\tau/2$. In ({\bf b}) and ({\bf c}), the region in between the yellow lines contains strong random perturbations in the underlying potential. ({\bf b}) Classical stable trajectories against perturbations. The black trajectory is a representative stable classical trajectory in the unperturbed potential, while the red trajectory which starts from o and ends at $e_{p}$ shows how the original stable trajectory is distorted by the perturbation. ({\bf c}) PCSPs against perturbations.  The black path connecting o and e is one of the PCSPs connecting o and e in the unperturbed potential, while the rest shows situations where stable paths in the perturbed potential merge into the original stable path. Note that once they spatially touch the PCSPs, it can no longer escape in the long range, which is very different from what one would expect from classical trajectories, where two trajectories can meet in space, but will drifit apart if there is a momentum mismatch. ({\bf d}) The black path from i to f via m is one PCSP consisting of two classical trajectories. The red and blue paths are the neighboring Feynman Paths that are in phase with each piece of the PCSP. The number of blue paths and red paths are intentionally chosen to be different to emphasize the fact that interference happens piecewisely.  }}
\label{dyna}
\end{figure*}

{\bf In the following, we shall assume that N is large enough, so (\ref{fp}) agrees with the quantum propagator. In other words, we are working in the small time step limit where the semiclassical propagator is known to converge to the quantum propagator\cite{b13}. }The summation is now over all possible piecewise classical trajectories(PCTs) connecting $(\vec{q}_{0},t_{0})$ and $(\vec{q}_{f},t_{f})$. The difference between (\ref{vvg}) and (\ref{fp}) is subtle, yet important in chaotic systems. In (\ref{vvg}), we only need to specify an initial momentum at $(\vec{q}_{0},t_{0})$ to uniquely determine the classical trajectory.  For PCTs in (\ref{fp}), we specify a $\vec{p}_{k}^{s}$ at any given $t_{k}$ and the classical equation of motion will yield a momentum $\vec{p}_{k+1}^{e}$ at $t_{k+1}$. Nonetheless, $\vec{p}_{k+1}^{e}$ is not necessarily the initial momentum at $t_{k+1}$. Instead, the trajectory from $t_{k+1}$ to $t_{k+2}$ can restart with any momentum $\vec{p}_{k+1}^{s}$ .  This difference is better illustrated with a figure, as in Fig.\ref{dyna}(a). Whereas a classical trajectory connecting A and C in Fig.\ref{dyna}(a) is uniquely specified by the initial momentum $\vec{p'}_A$ at A, both the initial momenta at A($\vec{p}_A$) and B($\vec{p}_B$) and the position of B($\vec{r}_B$) are needed for a PCT. To find the most stable classical trajectory, one needs to optimize with regards to only $\vec{p'}_A$. However, in order to find the most stable PCT, one needs to optimize with regards to $\vec{p}_A$, $\vec{p}_B$ and $\vec{r}_B$. Therefore, the most stable PCT is at least as stable as the most stable classical trajectory. In chaotic systems, all classical trajectories are unstable in the long range, {\bf where the long range is defined as large compared with the characteristic length leading to the chaos}. On the contrary, PCT has the ability to adjust its momentum at any point to follow the most stable paths. As we add more intermediate points between A and C,  the prefactor $\prod_{k} \left |\frac{\partial \vec{q}_{k+1}}{\partial \vec{p}_{k}^{s}}\right |^{-\frac{1}{2}}$ for an optimized PCT is exponentially larger than that of the most stable classical trajectory. Consequently, the long-range quantum dynamics is dictated by the least chaotic PCTs, which we name piecewise classical stable paths(PCSPs).

It is interesting to ask how PCSPs arise in such systems and the answer lies in the prefactor, which appears as a result of the stationary phase approximation used to derive VVG\cite{b13}. This prefactor summarizes the effect of the constructive interference of neighboring in-phase Feynman Paths and the interference is piecewise. PCSPs arise when the electrons follow a path that consists of many short paths that are continuously boosted by piecewise constructive interference. Consider the simplest case in Figure \ref{dyna}d, where the black path is part of one PCSP consisting of two short classical trajectories, one from i to m($pa_{i,m}$) and the other from m to f($pa_{m,f}$). In this case, $\left |\frac{\partial \vec{q}_{m}}{\partial \vec{p}_{i}}\right |^{-\frac{1}{2}}$ includes the contributions from all the red paths that are in phase with $pa_{i,m}$, while $\left |\frac{\partial \vec{q}_{f}}{\partial \vec{p}_{m}}\right |^{-\frac{1}{2}}$ includes the contributions from all the blue paths that are in phase with $pa_{m,f}$. The numbers of blue and red paths are intentionally chosen to be different to emphasize that the interference is piecewise. In classical mechanics, each classical trajectory corresponds to a point in the classical phase space spanned by ($\vec{p},\vec{q}$) at any given time. On the contrary, any quantum initial conditions must occupy a region with a finite volume in the phase space due to the uncertainty principle. Quantum dynamics can be modeled based on classical dynamics by assuming each classical trajectory carries a phase. Classical chaos implies that points starting close in phase space will separate from each other exponentially fast in the long range, which is due to the fact that any small deviation in initial conditions will be exponentially magnified over time. As a result, most classical trajectories will make negligible contributions due to their fluctuating phases. However, in the short range, certain regions in phase space, called stable regions\cite{d4} , will separate from each other relatively slower compared to the others purely by chance.  Points in these stable regions will carry similar phases and will contribute largely to the quantum propagator due to constructive interference. Since quantum dynamics can restart with any momentum at any time, paths consisting of a series of piecewise paths from stable regions will have dominant effect in the long range since by construction, they will have the smallest Lyapunov exponent\cite{d4}. 
\begin{figure*}
\begin{center}
\includegraphics[width=14cm]{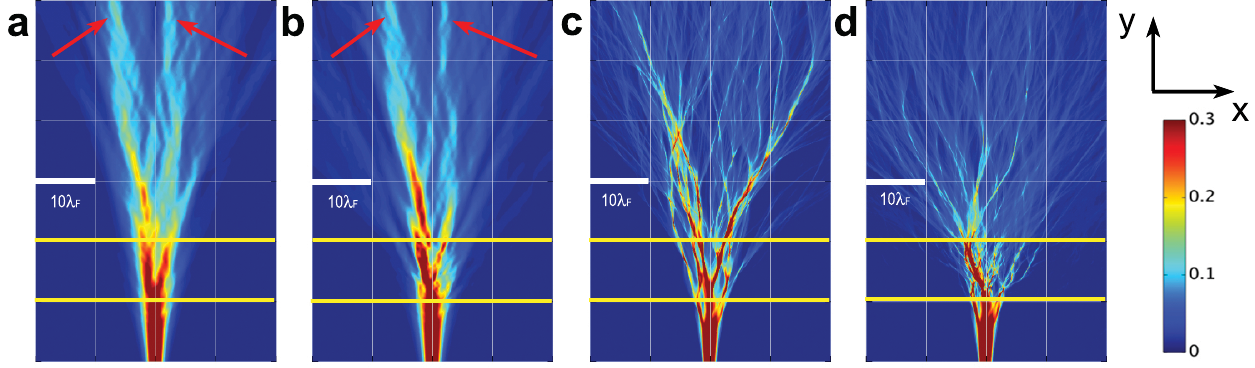}
\end{center}
\caption{(Color Online) \label{flux}\footnotesize{\textbf{Quantum Flux v.s. Classical Flux Patterns Along the y Direction}  The quantum and classical flux patterns over the unperturbed random potential are shown in ({\bf a}) and ({\bf c}) respectively. ({\bf b}) and ({\bf d}) show the respective quantum and classical flux patterns after the same perturbations are added to the regions between the yellow lines. The white reference grid denotes the same location in all images and all images starts at 15$\lambda_F$ from the injection point. The color axis shows the flux density per wavelength. As we can see, the same branches pointed to by the red arrows remain in ({\bf a}) and ({\bf b}), while all strong branches in ({\bf c}) are destroyed by the perturbations in ({\bf d}).More information about our methods can be found in the supplementary material.  }}
\end{figure*}

One property of such PCSPs that differs significantly from classical trajectories is its stability against perturbations. Unlike classical chaotic trajectories which are exponentially unstable to small perturbations,  PCSPs can actually tolerate a moderate amount of perturbations. In Fig.\ref{dyna}(c), the black path starting at o ending at e is one PCSP in the absence of perturbations.  After perturbations are added to the systems, the red path is one possible way how such PCSP can get recovered and the green and blue paths are two other possible paths that could help recover the original black path. The reason for such recovery lies in PCSPs' ability to adjust its momentum at any point. Once the perturbed path intersects the original path in the coordinate space, the original path can get recovered. This is very different from classical trajectories. Two classical trajectories with different momentum can intersect in coordinate space, but they will never be merged into the same path due to the difference in momentum. As a matter of fact, two such classical trajectories will separate from each other exponentially fast over time after intersection if the system is chaotic.

To make this difference more concrete, we provide an example where long range stability was not expected from existing theories, but should exist due to the arguments above. We consider the electron flow through a random potential in two dimensions. This system has been intensively studied in the context of Anderson Localization\cite{c1} and Universal Conductance Fluctuations\cite{c2}. Here, we consider a different regime where the random potential is weakly correlated and chaotic scattering results in the so-called "branched flow"\cite{b2}. Previously, it has been shown that classical trajectories combined with semiclassical initial conditions are enough to explain all the observed effects related to branched flow\cite{b10,d1,d2} and the branched flow pattern simulated using this approach bears close resemblance to the quantum flow pattern\cite{b2,d3,d4}. However, classical trajectories are exponentially unstable to perturbations and the prevailing theory on branching would imply that the branched flow pattern should be sensitive to perturbations. We numerically simulate both the quantum flow patterns and the branched flow pattern using classical trajectories with semiclassical conditions after an area of strong perturbations is introduced into the systems. More information about our numerical methods can be found in the supplementary material. The results are shown in Fig.1. All parameters(Fermi energy $E_{F}$, wavelength $\lambda_{F}$, donor to two-dimensional electron gases(2DEGs) distance, sample mobility etc.) are chosen to match that in a previous experiment\cite{b9}. The random potential has a standard deviation of $8\%E_{F}$ and correlation length $0.9\lambda_{F}$, as estimated in that experiment. The region of strong perturbations is introduced at 25$\lambda_{F}$ from the injection point and lasts for 10$\lambda_{F}$. In this region, a random potential of twice the original standard deviation($16\%E_{F}$) and the same correlation length is superimposed on the original random potential. As shown in Fig.1c and Fig.1d, all the branches involving only classical trajectories are destroyed in the long range(long in terms of the correlation length of the random potential). However, for the quantum simulations in Fig.1a and Fig.1b, the branches are only distorted inside and near the region of perturbations and recover themselves in the long range. This observation shall not be confused with that of a previous experiment\cite{b9}, in which case, classical theory suffices to explain the observed stability\cite{b10,b16}.

In the following, we show how PCSPs can explain this stability.  As shown in \cite{b15}, one can reproduce the experimental flow pattern by carefully constructing an initial wavepacket and propagating it through the scattering region. For such a wavepacket $\Psi(\vec{q},t_{0})$, it is given by $\Psi(\vec{q},t_{f})=\int d\vec{q}_{0}\ G(\vec{q},t_{f};\vec{q}_{0},t_{0})\Psi(\vec{q}_{0},t_{0})$ at $t_{f}$. Since electrons flow along narrow branches, $\Psi(\vec{q},t_{f})$ can be written as
\begin{equation} 
\Psi(\vec{q},t_{f})=\sum_{l} \Psi_{l}(\vec{q}) +\Psi_{r}(\vec{q}_{f})
\label{wp} \end{equation}
where each $\Psi_{l}(\vec{q})$ is a compact wavepacket corresponding to a branch and $\Psi_{r}(\vec{q})$ is a small residue. Time reversal symmetry implies that 
\begin{equation} 
\begin{aligned}
\Psi^{*}(\vec{q},t_{0})&=e^{-iH(t_{f}-t_{0})/\hbar} \Psi^{*}(\vec{q},t_{f})\\
&=\sum_{l} e^{-iH(t_{f}-t_{0})/\hbar} \Psi_{l}^{*}(\vec{q}) +e^{-iH(t_{f}-t_{0})/\hbar} \Psi_{r}^{*}(\vec{q})
\end{aligned}
\label{tr} \end{equation}
 
 $\Psi_{l}^{*}(\vec{q}) $ is nothing but a compact wavepacket and $ e^{-iH(t_{f}-t_{0})/\hbar} \Psi_{l}^{*}(\vec{q}) $ is the resulting wavepacket after $t_{f}-t_{0}$, which should consist of a new set of compact wavepackets corresponding to the resulting branches that can be written as

\begin{equation} 
 e^{-iH(t_{f}-t_{0})/\hbar} \Psi_{l}^{*}(\vec{q}) =\sum_{l_{s}}\Psi_{l,l_{s}}^{*}(\vec{q})+\Psi_{l,r}^{*}(\vec{q})
\label{bb} \end{equation}
Plugging (\ref{bb}) into (\ref{tr}),
\begin{equation} 
\Psi^{*}(\vec{q},t_{0})=\sum_{l,l_{s}}\Psi_{l,l_{s}}^{*}(\vec{q})+\sum_{l} \Psi_{l,r}^{*}(\vec{q})+e^{-iH(t_{f}-t_{0})/\hbar} \Psi_{r}^{*}(\vec{q})
\label{trr} \end{equation}

$\Psi^{*}(\vec{q},t_{0})$ consists of one or several compact wavepacket at typical experimental temperature\cite{b15}. If $\Psi_{l,l_{s}}^{*}(\vec{q})$ barely overlaps with $\Psi^{*}(\vec{q},t_{0})$ in space, its contribution has to be cancelled by other wavepackets and (\ref{trr}) can be further simplified to 

\begin{equation} 
\Psi(\vec{q},t_{0})=\sum_{l,l_{s'}}\Psi_{l,l_{s'}}(\vec{q})+\Psi_{r'}(\vec{q})
\label{trrr} \end{equation}
where the summation is over wavepackets that spatially overlap with $\Psi(\vec{q},t_{0})$. 

Every PCSP is reversible,so the major contribution to each $\Psi_{l}(\vec{q})$ comes from $\Psi_{l,l_{s'}}(\vec{q})$, which, by construction, occupies a more localized region in space than $\Psi(\vec{q},t_{0})$. 

We now consider how the perturbed region influences each $\Psi_{l,l_{s'}}(\vec{q})$. First of all, we need to remember that the disorder is weak, resulting in only small angle scattering. If we consider the wavepacket($\Psi_{0,pert}(\vec{q})$) after passing through the perturbed region, it will spatially overlap largely with the one($\Psi_{0,unpert}(\vec{q})$) in the absence of perturbations. This can be inferred from our quantum simulations in Fig.1a and Fig.1b and the reason is that small angle scattering are not effective in reducing spatial overlap. However, it can reduces the coherence between the two wavepackets through phase randomization. One can model this phase randomization by changing the momentum of each $\Psi_{l,l_{s'}}(\vec{q})$. That is, we assume $\Psi_{l,l_{s'}}(\vec{q})\approx f_{0}(\vec{q}-\vec{q}_{l,0}) e^{i\vec{p}_{o}\cdot (\vec{q}-\vec{q}_{l,0})} $ and the pertubed region changes it to $\Psi_{l,l_{s'},p} \approx f_{0,p}(\vec{q}-\vec{q}_{l,0}) e^{i\vec{p}_{p}\cdot (\vec{q}-\vec{q}_{l,0})} $, where both $f_{0}(\vec{q}-\vec{q}_{l,0}) $ and $f_{0,p}(\vec{q}-\vec{q}_{l,0})$ are real and have large spatial overlap. In the absence of perturbations, assume that the resulting wavepacket at $t_f$ is $\Psi_{l}(\vec{q})\approx  f_{f}(\vec{q}-\vec{q}_{l,f})e^{i\vec{p}_{f}\cdot (\vec{q}-\vec{q}_{l,f})} $, with  $f_{f}(\vec{q}-\vec{q}_{l,f})$ real. The effect of the perturbed region is to change the initial momentum of each wavepacket corresponding to a branch. In free space, a change in the initial momentum will drift wavepackets apart linearly fast with time. \textbf{However,with PCSPs, the two wavepackets are actually pulled towards each other.}  To see the reason, we consider the following integral that measures the probability that the perturbed wavepacket will lead to the same original branch. 
\begin{equation} 
\begin{aligned}
&<\Psi_{l}|e^{-iHt/\hbar}|\Psi_{l,l_{s'},p}>= \int d\vec{q}_{0} d\vec{q}_{f} \ \Psi_{l,l_{s'},p}(\vec{q}_{0})\Psi_{l}^{*}(\vec{q}_{f})\\
&\times \frac{1}{(2i\pi\hbar)^{N}}\sum_{j} \prod_{k} \left |\frac{\partial \vec{q}_{k+1}}{\partial \vec{p}_{k}}\right |^{-\frac{1}{2}} e^{iS_{j}(\vec{q}_{k+1},\vec{q}_{k},t_{k+1},t_{k})/\hbar -i\frac{\pi}{2}v_{j,k,k+1}}
\end{aligned}
\label{inte} \end{equation}

This is by no means a simple integral. Luckily, the prefactor $\prod_{k} \left |\frac{\partial \vec{q}_{k+1}}{\partial \vec{p}_{k}}\right |^{-\frac{1}{2}}$ ensures that we only need to consider the neighborhood of PCSPs when $\vec{q}_{l,0}$ and $\vec{q}_{l,f}$ are far apart. Moreover, the integrand is oscillatory, so we focus on stationary phase points, the conditions for which are\cite{b18}

\begin{equation} 
\frac{\partial S_{j}}{\partial \vec{q}_{o}}+\vec{p}_{o}=0\,\,\,\,\,\,\,\,\,\,\,\,\frac{\partial S_{j}}{\partial \vec{q}_{f}}-\vec{p}_{f}=0
\label{sp} \end{equation}


For PCSPs, Equation (\ref{sp}) is equivalent to
\begin{equation} 
\sum_{k=1}^{N-1} (\vec{p}_{k}^{e}-\vec{p}_{k}^{s}) \cdot \frac{\partial \vec{q}_{k}}{\partial \vec{q}_{o}}+(\vec{p}_{o}-\vec{p}_{o}^{s})=0\\
\sum_{k=1}^{N-1} (\vec{p}_{k}^{e}-\vec{p}_{k}^{s}) \cdot \frac{\partial \vec{q}_{k}}{\partial \vec{q}_{f}}+(\vec{p}_{f}^{e}-\vec{p}_{f})=0
\label{stable path} \end{equation}

In the case of a single classical trajectory connecting $\vec{q}_{f}$ and $\vec{q}_{o}$, (\ref{stable path}) yields $\vec{p}_{f}^{e}=\vec{p}_{f}$ and $\vec{p}_{o}=\vec{p}_{o}^{s}$, which is the classical limit as $\hbar \to 0$. 

In order to accommodate a moderate change in $\vec{p}_{o}$, we can make small adjustments to each $\vec{p}_{k}^{s}$ such that (\ref{stable path}) still holds. Mathematically, it is given by
\begin{equation} 
\delta \vec{p}_{o}=\sum_{k=0}^{N-1} (\delta\vec{p}_{k}^{s}-\sum_{i>k} (\frac{\partial \vec{p}_{i}^{e} }{\partial\vec{p}_{k}^{s}}\delta \vec{p}_{k}^{s}\cdot \frac{\partial \vec{q}_{i}}{\partial \vec{q}_{o}}
-(\vec{p}_{i}^{s}-\vec{p}_{i}^{e}) \cdot \frac{\partial^{2} \vec{q}_{i}}{\partial \vec{q}_{o}\partial \vec{p}_{k}^{s}}\delta\vec{p}_{k}^{s}))
\label{stablepath} \end{equation}

The above argument shows the existence of stationary phase points in the vicinity of PCSPs if $\left | \vec{q}_f-\vec{q}_o\right |$ is long compared to the correlation length of the random potential causing the chaos. This implies that even if the two wavepackets have different initial momenta, they can still end up in the same branch. In the example provided above, the simulated region is about seventy correlation lengths, which is considered to be long range in this case.

In summary, we've shown that interference is of piecewise nature in open quantum systems where the classical dynamics is chaotic. This piecewise nature gives rise to a long range stability in branched flow that challenges the prevailing interpretation of branching. Our theory is an extension to the traditional theory of scattering and the result on the long range stability of branched flow can be tested experimentally in either 2DEGs or photonic systems\cite{b17}. Moreover, our theory shall have implications for the wave dynamics in many other open system where the classical ray dynamics is chaotic.

I thank Prof. Eric J. Heller for many helpful discussions and financial support from the U.S. Department of Energy under DE-FG02-08ER46513.


\begin{thebibliography}{99}
\bibitem{b16}
In that experiment, the QPC is shifted by about one correlation length in an attempt to create a big change in initial conditions. However, as explained previously\cite{b10}, even if the two initials wavepackets are almost nonoverlapping, the adiabaticity of the QPC would still correlate the two wavepackets afterwards. This creates large overlap in phase space when the electrons get away from the QPCs, which is essentially the reason leading to the observed stability in that experiment\cite{b9}. In our case, we create the region of perturbations after electrons pass through the QPC, so no overlap is produced by the QPC. Moreover, our perturbations are strong enough to destroy the overlap in phase space. The easiest way to see the difference is by noting that classical stable branches are destroyed in our case, but classical trajectories alone can reproduce the stability in that experiment\cite{b10}.
\bibitem{b18}  
We are also implicitly assuming that no focal point is missed in the neighborhood of the PCSPs, which is a good approximation since the condition for focal points is $\left |\frac{\partial \vec{q}_{f}}{\partial \vec{p}_{0}}\right |=0$. In the vicinity of a PCSP, we shall expect the number of focal points to be fixed.
\end{thebibliography}
\end{document}